\documentclass[%
 reprint,
superscriptaddress,
 amsmath,amssymb,
 aps,
prb,
]{revtex4-2}

\usepackage{graphicx}
\usepackage{dcolumn}
\usepackage{bm}

\usepackage{orcidlink}

\begin{document}

\preprint{APS/123-QED}

\title{High-efficiency broadband achromatic metalens in the visible}

\author{Liang Hou}
\affiliation{School of Physics and Materials Science, Nanchang University, Nanchang 330031, China}

\author{Hongyuan Zhou}%
\affiliation{School of Physics and Materials Science, Nanchang University, Nanchang 330031, China}

\author{Dandan Zhang}%
\affiliation{School of Physics and Materials Science, Nanchang University, Nanchang 330031, China}

\author{Ganqing Lu}%
\affiliation{School of Physics and Materials Science, Nanchang University, Nanchang 330031, China}

\author{Dejiang Zhang}%
\affiliation{School of Physics and Materials Science, Nanchang University, Nanchang 330031, China}

\author{Tingting Liu~\orcidlink{0000-0003-0577-6310}}%
 \email{ttliu@ncu.edu.cn}
  \affiliation{School of Information Engineering,
Nanchang University, Nanchang 330031, China}
 \affiliation{Institute for Advanced Study, Nanchang University, Nanchang 330031, China}
 
\author{Shuyuan Xiao~\orcidlink{0000-0002-4446-6967}}%
 \email{syxiao@ncu.edu.cn}
 \affiliation{School of Information Engineering,
Nanchang University, Nanchang 330031, China}
 \affiliation{Institute for Advanced Study, Nanchang University, Nanchang 330031, China}

\author{Tianbao Yu~\orcidlink{0000-0003-4308-7022}}%
 \email{yutianbao@ncu.edu.cn}
\affiliation{School of Physics and Materials Science, Nanchang University, Nanchang 330031, China}%

\date{\today}

\begin{abstract}
The metalenses have been extensively studied for their compact and flexible characteristics in focusing and imaging applications. However, it remains a significant challenge to design a broadband achromatic metalens that maintains high efficiency under arbitrary polarization incidence. In this work, we design a broadband achromatic metalens that achieves polarization-independent, high-efficiency focusing by effectively utilizing both co-polarization and cross-polarization components of the transmitted light. Using a minimalist anisotropic nanofin library, we optimize the phase distribution of the metalens at each designed wavelength with the particle swarm algorithm. Numerical simulations demonstrate a stable focal length with a deviation of less than 4$\%$ and an average focusing efficiency of 80.5$\%$ in the visible wavelength range of 450 to 650 nm. Moreover, we design a multi-wavelength off-axis bi-focal metalens to demonstrate the flexible control of output light phase and dispersion achieved by this method. The generality of this design enables its implementation in various metasurface devices, accelerating applications in broadband imaging and virtual/augmented reality.

\end{abstract}

\maketitle

\section{Introduction}
Metasurfaces are two-dimensional planar structures composed of engineered subwavelength electromagnetic resonators. Through the strategic arrangement of metaunits, the coupling between electromagnetic radiation and matter can be substantially amplified, enabling the agile modulation of diverse light characteristics, including amplitude, phase, polarization, and frequency \cite{1,2,3,4,5,6,7,8,A1,A8}. Benefiting from precise wavefront manipulation, metasurfaces have enabled novel advancements in designing modern optical components, such as beam deflectors \cite{9,10,11,A7}, metalenses \cite{12,13,14,15,16,17,18,19,20,21,22,23,A2,A3,A6}, metasensors \cite{24,25,26}, metaholograms \cite{27,28,29,30,31,32,33,34,A4,A5}, and nanoprinted imaging systems \cite{35,36,37}. However, a significant challenge in applying metasurfaces to imaging systems requiring broadband or multi-wavelength operation is their inherent chromatic aberration \cite{38}, akin to traditional diffractive devices. To address this issue, various design schemes of metalens have been proposed, including multilayer metasurfaces \cite{39,40,41,42,43}, spatial multiplexing \cite{44,45}, and zone interference \cite{46,47,48}. While these approaches can achieve consistent focal lengths across multiple wavelengths, they often involve trade-offs, such as reduced focusing efficiency, complex manufacturing processes, limited structural freedom, and challenges related to interlayer coupling and alignment.

Metalenses utilizing the Pancharatnam-Berry (PB) phase have partially mitigated the limitations associated with conventional metasurface designs \cite{44,49,50}. Recent research has demonstrated the use of single-layer nanostructures to achieve diffraction-limited achromatic focusing and imaging across a wide visible spectrum, from 470 to 670 nm. However, the PB phase modulation approach necessitates circularly polarized incident light, thereby increasing the complexity of the optical system and inherently limiting the theoretical efficiency less than 50$\%$. This challenge can be overcome by using symmetrical cylindrical or square nanopillars as the constituent structures \cite{38,44,51}. While these symmetric designs can provide polarization insensitivity, they also constrain the design space and flexibility, potentially limiting their applicability in high-quality imaging systems. An alternative approach is to restrict the rotation angle of each anisotropic structure to 0 or 90 degrees, thereby increasing design freedom while maintaining polarization insensitivity. This method has been widely adopted in metaholograms \cite{52} and metalens \cite{53,54} design. Some of the latest works have achieved achromatic and polarization-insensitive performance in the visible \cite{55} and infrared \cite{56}, while maintaining diffraction-limited performance. However, previous PB phase-based designs have often prioritized polarization conversion terms while treating the conservation terms as stray light, resulting in energy loss and reduced focusing efficiency across a wide bandwidth. When designing a metalens, it is imperative to meticulously manage the contributions of both polarization conversion and conservation terms to maximize the device's overall performance and efficiency.

In this study, we employ a minimalistic anisotropic TiO$_{2}$ nanofin as a metaunit to compile a phase response database encompassing polarization conversion and conservation terms of transmitted light. This database enables us to characterize various phase profiles and dispersions. Based on the asymptotic phase compensation method, we optimize the phase profile of the metalens using a particle swarm optimization algorithm to minimize phase matching errors. We demonstrate an achromatic metalens capable of accommodating arbitrary polarization incidence, depicted in Fig.~\ref{Figure1}(a). Over the wavelength range from $\lambda$=450 to 650 nm, the maximum error in focal length displacement is merely 4$\%$, with diffraction-limited focal spots. Taking full advantage of both polarization conversion and conservation terms for focusing, the average focusing efficiency reaches 80.5$\%$. To further demonstrate the universality of this design, we also develop a multi-wavelength off-axis bi-focal metalens capable of converging the two polarization states of transmitted light to distinct focal points. This showcases the potential applications of our design in virtual reality (VR) and augmented reality (AR) and color display technologies.

\section{Principle of high-efficiency broadband achromatic focusing}
When light passes through an anisotropic nanofin, the transmitted electric field can be described by the Jones vector \cite{57},
\begin{equation}
\left(
\begin{matrix}
\widetilde{E}_{x} \\
\widetilde{E}_{y}
\end{matrix}
\right)= \frac{\widetilde{t}_l+\widetilde{t}_s}{2}\left( 
\begin{matrix}
1 \\
\pm i    
\end{matrix}
\right)+\frac{\widetilde{t}_l-\widetilde{t}_s}{2}{e^{\pm i2\alpha}}\left(
\begin{matrix}
1 \\
\mp i    
\end{matrix}
\right),
\label{eq 1}
\end{equation}
where $\widetilde{t}_{l}$ and $\widetilde{t}_{s}$ are frequency-dependent complex transmission coefficients when incident light is linearly polarized along its short and long axes, respectively. The first and second terms are called polarization conserved and converted terms, respectively. And then, their efficiencies can be expressed as 
\begin{equation}
T_{co}= {\lvert \frac{\widetilde{t}_l+\widetilde{t}_s}{2} \rvert}^2, 
T_{cross}= {\lvert \frac{\widetilde{t}_l-\widetilde{t}_s}{2} \rvert}^2.
\label{eq 2}
\end{equation}

Meanwhile, the phase of co-polarization and cross-polarization are  $\phi_{co}$=$arg(\frac{\widetilde{t}_l+\widetilde{t}_s}{2})$ and $\phi_{cross}=arg(\frac{\widetilde{t}_l-\widetilde{t}_s}{2})$, respectively. The term $e^{\pm i2\alpha}$ in the polarization conversion mechanism elucidates the genesis of the PB phase, often resulting in polarization sensitivity due to distinct values acquired for left-circularly polarized (LCP) and right-circularly polarized (RCP) incident light, namely $e^{i2\alpha}$ and $e^{-i2\alpha}$. To mitigate the impact of this PB phase on the cross-polarization phase, nanofins can be configured at orientations where $\alpha$ equals 0° or 90°.

\begin{figure}
\includegraphics[width=\linewidth]{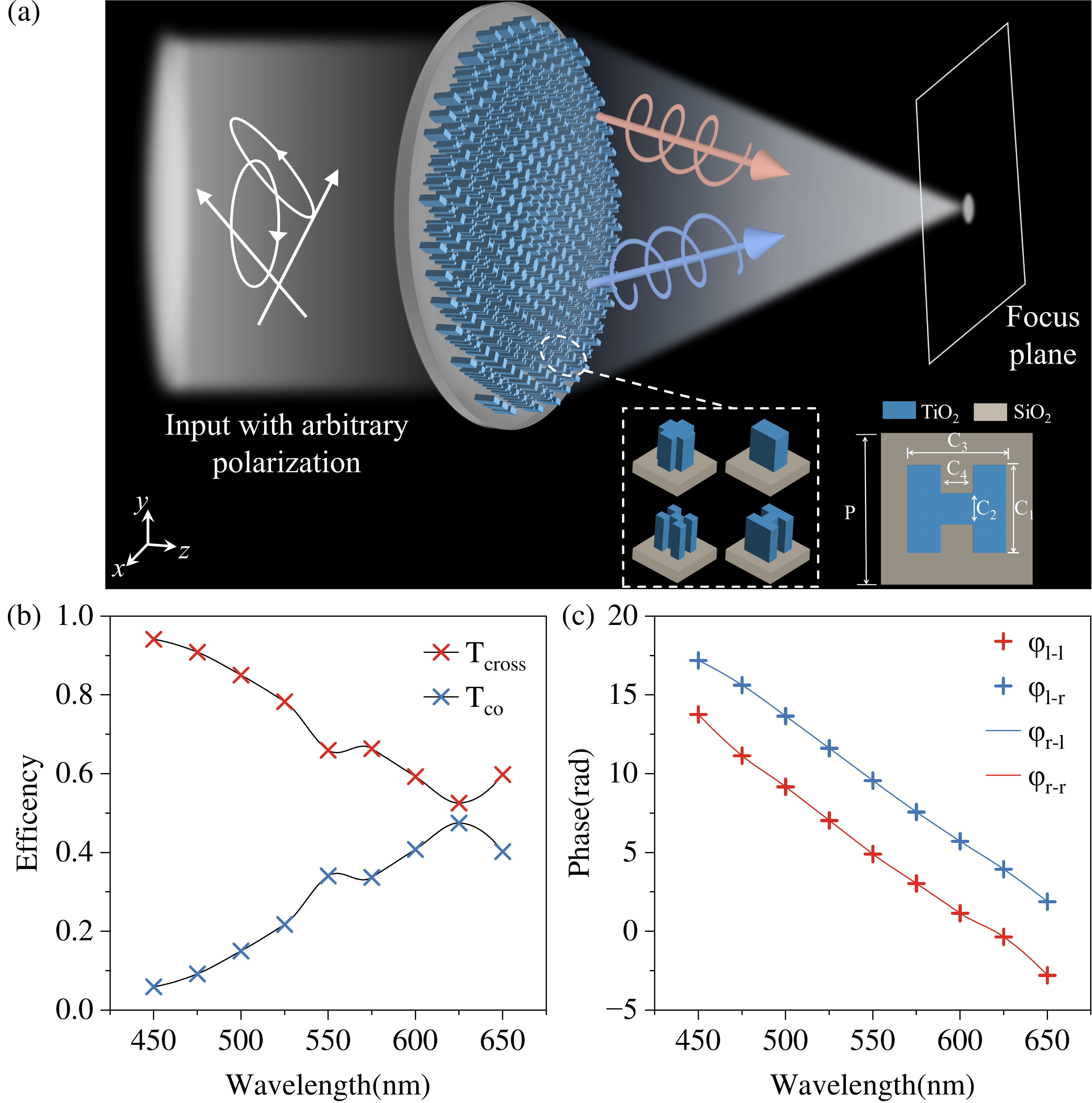}
\caption{\label{Figure1}The schematic diagram of the achromatic metalens for the principle of polarization insensitivity and high efficiency. (a) The schematic diagram of anisotropic metalens which can focus the polarization conversion and conservation component on the same point. The metalens is composed of four basic structures, characterized by parameters $C_{1},C_{2},C_{3}$ and $C_{4}$, which range from 60 to 340 nm. (b) The variation of polarization conversion and conservation efficiency of the nanofin as a function of wavelength. (c) The phase shift of the transmitted electric field component under two circularly polarized lights. The legend $\phi_{l-r}$ indicates the phase of RCP transmitted light under LCP incidence. Blue and red colors represent cross-polarization and co-polarization, respectively.}
\end{figure}

Fig.~\ref{Figure1}(b) illustrates the variation of the polarization conversion and polarization conservation efficiency of the nanofin across different wavelengths. It is known that the sum of these two efficiencies remains constant, expressed as $T_{co}+T_{cross}=1$. However, previous studies have traditionally regarded the conservation term as stray light. In specific wavelength bands, notably $\lambda > 525$ nm, disregarding the conservation term can lead to energy loss and diminished focusing efficiency across a broad spectrum. To address this issue, we adjust $\phi_{co}$ and $\phi_{cross}$ to align with the phase requirements of the two polarization components. This alignment ensures that transmitted co-polarization and cross-polarization light converge precisely at the same focal point. Theoretically, this approach aims to optimize the utilization of transmitted light energy, thereby achieving highly efficient focusing. Furthermore, under both LCP and RCP incidences, the co-polarization and cross-polarization phases remain unchanged, as depicted in Fig.~\ref{Figure1}(c). This characteristic implies that $\phi_{l-l} = \phi_{r-r}$ and $\phi_{l-r} = \phi_{r-l}$, affirming that the metalens exhibits polarization insensitivity and can effectively focus light of arbitrary polarization \cite{58}. Consequently, highly efficient and polarization-insensitive metasurface devices can be achieved.

\begin{figure}
\includegraphics[width=\linewidth]{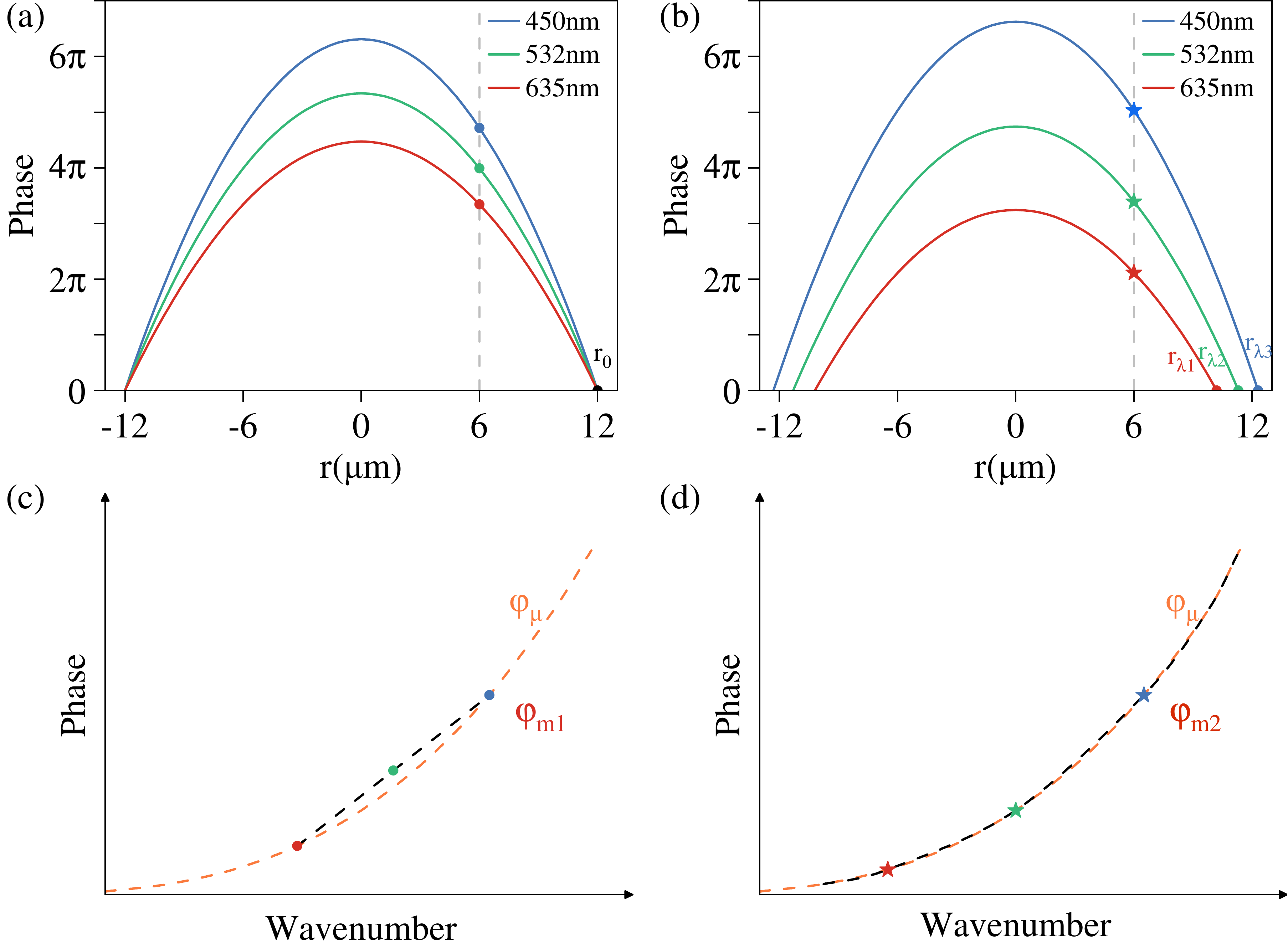}
\caption{\label{Figure2}The principle of linear dispersion in (a,c) and asymptotic dispersion in (b,d). (a) The traditional achromatic metalens scheme. (b) The optimized phase profile diagram of the achromatic metalens scheme. The comparison between the intrinsic phase dispersion of the metaunit and the constructed phase dispersion using (c) the linear phase and (d) asymptotic phase compensation methods. The three different wavelengths are represented by red, green, and blue, respectively, with solid circles indicating linear scattering and stars representing gradual scattering.}
\end{figure}

To achieve achromatic focusing, the metalens should provide independent phase profiles for each design wavelength. This necessitates that each anisotropic nanofin within the metalens provides specific cross-polarization and co-polarization phase responses tailored to its position, ensuring alignment with the phase requirements across all design wavelengths. For a nanofin of specific shape, the phase response via the waveguide effect is given by $\phi_{u}=kn_{eff}H$ \cite{51}, where $n_{eff}$ represents the effective refractive index associated with the cross-sectional geometry of the nanofin, $k$ denotes the wavenumber, and $H$ signifies the height of the nanofin. Indicated by the orange dotted line in Fig.~\ref{Figure2}(c) and ~\ref{Figure2}(d), the phase response exhibits a positive correlation with the wavenumber $k$. Additionally, the wavelength dependence of $n_{eff}$ results in a nonlinear relationship.

For a conventional converging spherical metalens with focal length $f$, its phase profile has to follow \cite{38},
\begin{equation}
\phi_{m1}= -\frac{2\pi}{\lambda}(\sqrt{r^2+f^2}-\sqrt{{r_{0}}^2+f^2}),
\label{eq 3}
\end{equation}
where $\lambda$ is the design wavelength, $r$ is the radial distance from the center of the metalens, and $r_{0}$ is the radius of the metalens. The phase corresponding to $r_{0}$ is defined as the reference phase. Fig.~\ref{Figure2}(a) illustrates the spatial distribution of phase required for the metalens at three specified design wavelengths. The phase variation exhibits a positive correlation particularly at positions within $r_{0}$ and it is possible of the metaunit to effectively provide phase compensation. However, the inherent linear relationship between phase and wavenumber introduced by $r_{0}$ can only provide a linear approximation of the phase dispersion across the metaunit. As depicted in Fig.~\ref{Figure2}(c), wavelengths represented by the red and blue solid points satisfy the phase requirements, whereas those corresponding to the green points exhibit phase mismatches. Moreover, the constrained selection of metaunits in the database leads to significant phase matching errors.
The asymptotic phase compensation method \cite{23} enables more precise alignment with the intrinsic phase dispersion response of nanostructures. This approach involves adjusting the designed wavefronts at different wavelengths, thereby addressing the challenge of substantial errors that may arise when employing a linear phase compensation scheme to achieve arbitrary dispersion operations. Based on this asymptotic phase compensation approach, we successfully construct the phase profile of the metalens depicted in Fig.~\ref{Figure2}(b). This method facilitates meticulous control over the optical characteristics of the nanostructure by accommodating its wavelength-dependent phase response. The constructed phase profile can be expressed as
\begin{equation}
\phi_{m2}= -\frac{2\pi}{\lambda}(\sqrt{r^2+f^2}-\sqrt{{r_{\lambda}}^2+f^2}).
\label{eq 4}
\end{equation}

Here, $r_{\lambda}$ represents a parameter associated with wavelength, determined through the Particle Swarm Optimization (PSO) algorithm. When circular polarization is incident, both the polarization conversion term and the conservation term conform to their respective phase requirements. The discrepancy between the phase response of the nanofin and the phase distribution of the lens at N designed wavelengths is quantified by \cite{59}
\begin{equation}
Error= \sum_{i}^{L,R}\sum_{k=1}^{N}{\lvert \phi_{m2}^{i}(r_{\lambda}^{i},\lambda_{k})-\phi_{u}^{i}(\lambda_{k}) \rvert}.
\label{eq 5}
\end{equation}

By iteratively optimizing the particle swarm algorithm, the nanofin with the minimal matching error is selected to assemble the metalens. This approach ensures that the phase dispersion of the constructed metalens closely approximates that of the nanofin, as depicted in Fig.~\ref{Figure2}(d). Notably, the phase distribution at each designed wavelength exhibits distinct displacements at the same focal point. This capability allows for the construction of a metalens capable of performing arbitrary dispersion operations, precisely aligning the dispersion and phase of both the polarization conversion and conservation terms of the nanofins.

\section{Design and results of Broadband Achromatic Metalens}  

Our design of a polarization-insensitive and achromatic metalens begins with a parameter sweep of the element depicted in the inset of Fig.~\ref{Figure1}(a) to construct a metaunit library. The phase dispersion of the nanofins is strongly correlated with their cross-sectional shape, offering a significant degree of freedom for dispersion manipulation. Considering the reliability and simplicity of fabrication, our database comprises four basic structures, as shown in Fig.~\ref{Figure1}(a). The nanofin substrate is SiO$_{2}$, with a period $P=400$ nm. The upper structure is composed of TiO$_{2}$, a high-refractive-index dielectric material in the visible range, capable of achieving a broader range of dispersion at a lower nanometer scale.

\begin{figure*}
\includegraphics[width=\linewidth]{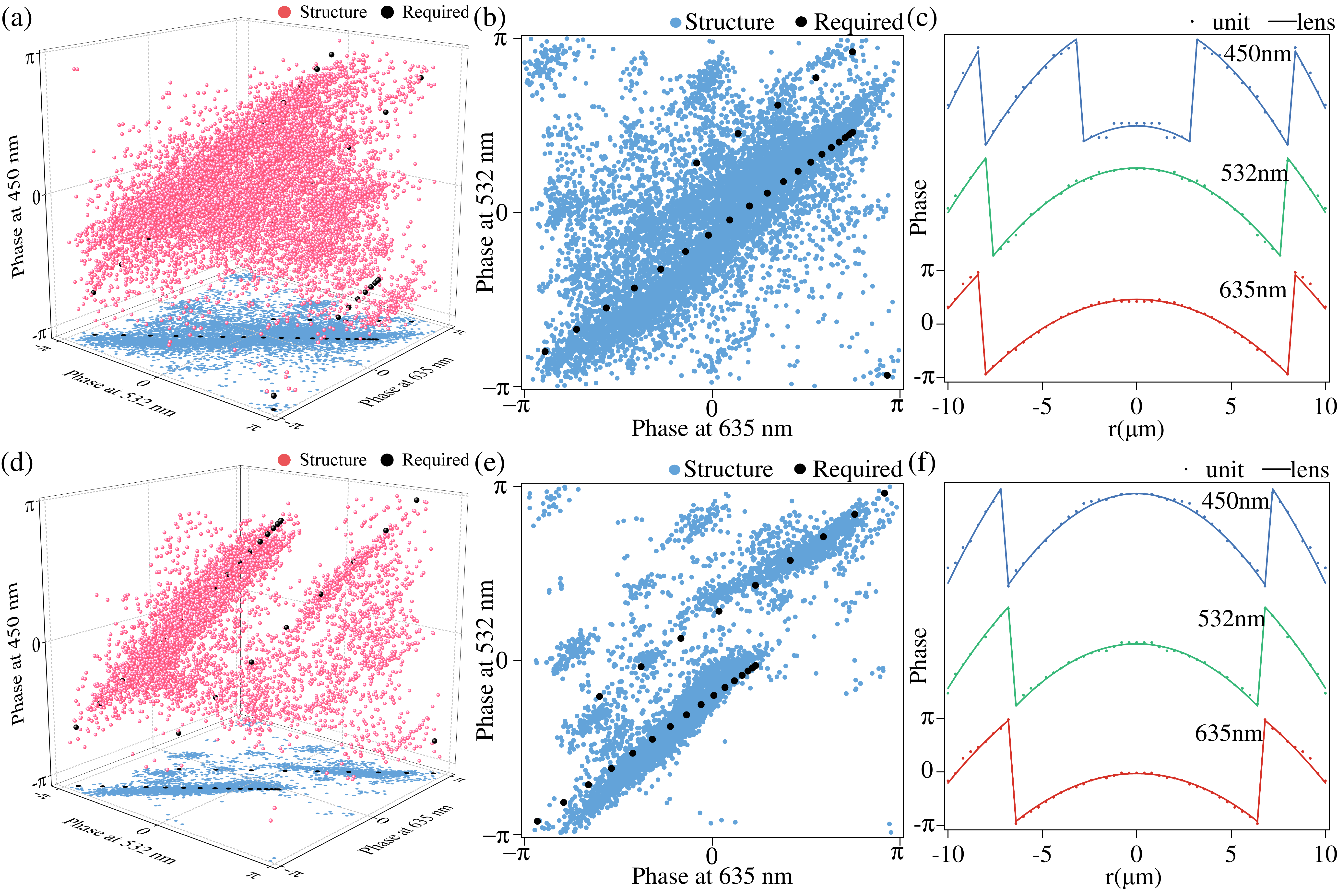}
\caption{\label{Figure3}The schematic matching results of structural phase and metalens phase. (a,d) The database of phase responses for the cross-polarization and co-polarization components of various nanofins under LCP incidence. Each red dot represents the phase response of a specifically designed nanofin, while the black dot represents the phase required at different positions on the metalens, with coordinates corresponding to the folded phase at 450, 532, and 635 nm. (b,e) The projection of the structural phase and metalens phase at coordinates 532 and 635 nm for the cross-polarization and co-polarization components, respectively. (c,f) The matching results for different wavelengths in the radial dimension of the metalens with a 10 $\mu$m radius for the cross-polarization and co-polarization components. The solid lines represent the constructed wavefront phase profiles, while the dots represent the phases of the matched nanofins.}
\end{figure*}

These four structures are derived through a parametric sweep using the finite-difference time-domain (FDTD) method, with $C_{1},C_{2},C_{3}$ and $C_{4}$ ranging from 60 to 340 nm, and a height $H$ of 600 nm. The FDTD solver is employed to determine the phases of both the polarization conversion and conservation term for each nanofin at three wavelengths: $\lambda=450$, 532, and 635 nm. This process generates a database of over 12,000 structures. To ensure reliability, we impose a minimum width constraint of 50 nm and a maximum aspect ratio of 12:1, and exclude structures exhibiting low efficiency or phase anomalies due to coupling effects. The refined structure library ultimately comprises more than 9,000 distinct nanofins. For detailed simulation details, please refer to Section 1 of the Supplementary Materials.
Fig.~\ref{Figure3}(a) and ~\ref{Figure3}(d) present the phase response databases for the polarization conversion term and the conservation term of transmitted light under LCP incident light, for nanofins with varying parameters. Each red dot indicates the phase response of a specifically designed nanostructure, while the black dot represents the phase required for the metalens, as determined by Eq.~(\ref{eq 3}). The coordinates of the red dot correspond to the folded realized phases at 450, 532, and 635 nm. To streamline the design process, we consider only three wavelengths. The data points collectively form a point cloud dispersed within the $2\pi\times2\pi\times2\pi$ phase space. Folding the phase values between $-\pi$ and $\pi$ simplifies the phase space, thereby circumventing the complexity associated with maintaining a one-to-one relationship in the unfolded phase space. Fig.~\ref{Figure3}(b) and ~\ref{Figure3}(e) display the projections of the phase provided by the nanofins and the phase required for the metalens focusing at the coordinates corresponding to 532 and 635 nm. By constructing the phase using Eq.~(\ref{eq 3}) and optimizing $r_{\lambda}$ for different wavelengths through the PSO algorithm, the nanofins with the minimal matching error are identified to construct a metalens. Fig.~\ref{Figure3}(c) and~\ref{Figure3}(f) illustrate the phase matching results for co-polarization and cross-polarization at three wavelengths and various radial distances for a metalens with a 10 $\mu$m radius under LCP. The solid line represents the constructed wavefront phase profile, and the dots denote the phases achieved by the corresponding matching nanofins. These selectd nanofins closely approximate the phase requirements of the achromatic metalens.

\begin{figure}
\includegraphics[width=\linewidth]{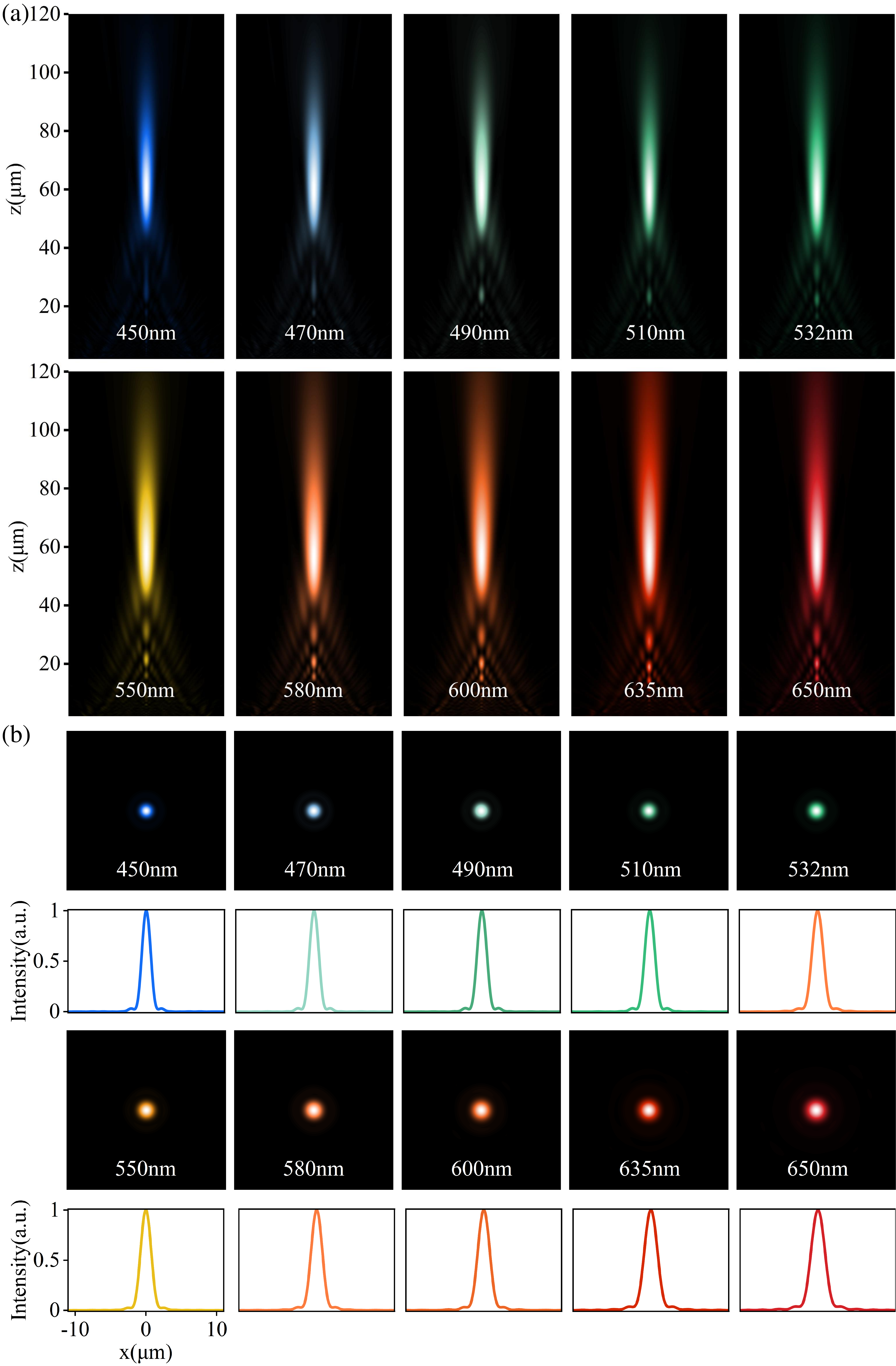}
\caption{\label{Figure4}Simulated focusing intensity distribution at different wavelengths. (a) The normalized intensity distribution on the \textit{x-z} plane is depicted in pseudo colors, corresponding to the respective wavelengths in the visible light spectrum from wavelength $\lambda$= 450 to 650 nm. (b) Normalized intensity distribution of the achromatic metalens on the \textit{x-y} plane.}
\end{figure}

Although the metalens is designed for three specific wavelengths, it operates effectively across a broad range within the visible spectrum. To demonstrate this, we design an achromatic metalens with a diameter of 20 $\mu$m and a numerical aperture (NA) of 0.164, as shown in Fig.~\ref{Figure1}(a). The achromatic performance of the designed metalens is investigated through simulations employing the FDTD method. Fig.~\ref{Figure4}(a) presents the normalized intensity distributions of the proposed achromatic metalens at various wavelengths. This metalens focuses almost the entire visible light spectrum at the same position. Fig.~\ref{Figure4}(b) illustrates the simulated focal spot intensity distribution along with the corresponding cross-section on the focal plane. The focal spots exhibit circular symmetry, characterized by an Airy disk distribution. For comparison, we also design a chromatic metalens with the same diameter and numerical aperture. The chromatic metalens is designed using rotated nanofins of the same length and width, as detailed in Section 2 of the Supplementary Materials. The chromatic metalens represents the case where there is no regulated dispersion and exhibits a focal length shift similar to that of a Fresnel lens. Supplementary Material Fig.~S4(e) presents the normalized intensity distributions of the proposed chromatic metalens at various wavelengths, highlighting significant fluctuations in its focal length that indicate pronounced chromatic aberration.

\begin{figure}
\includegraphics[width=\linewidth]{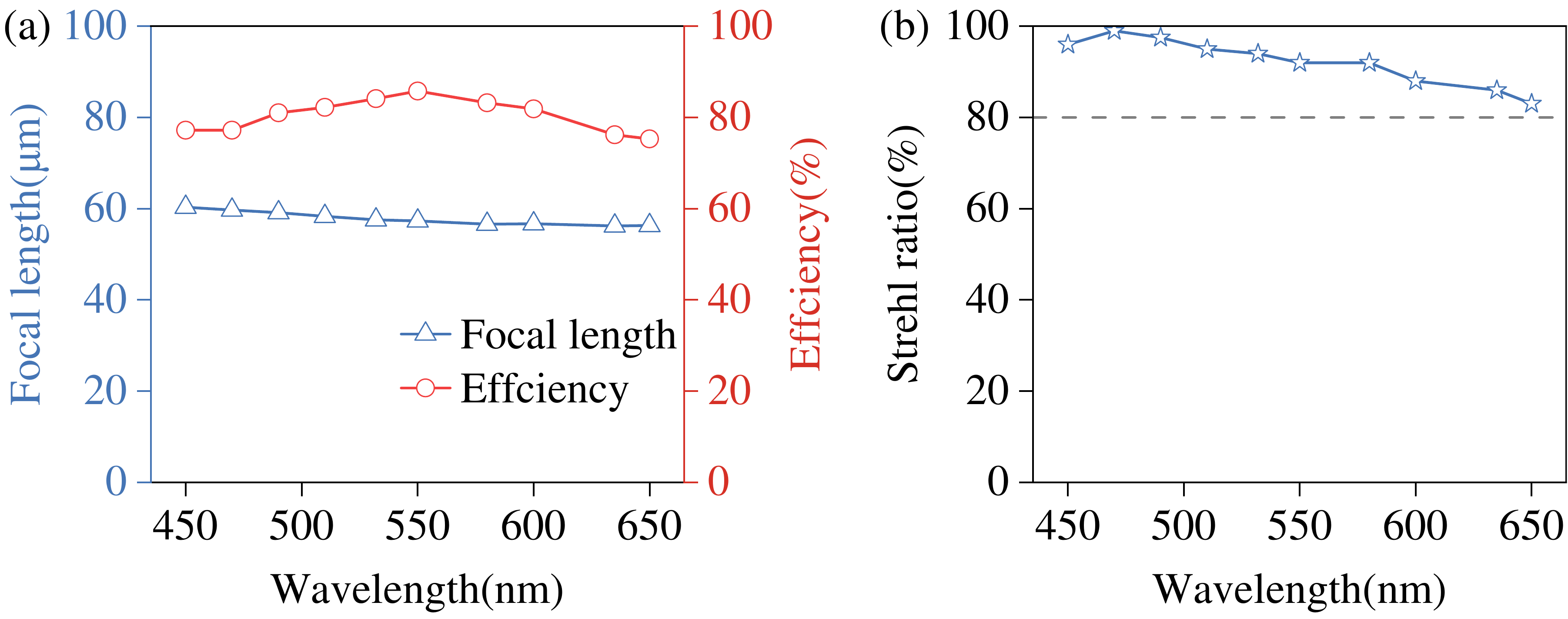}
\caption{\label{Figure5}Simulated focal lengths, focusing efficiencies and strehl ratio at different wavelengths. (a) Focal length and focusing efficiency of the metalens. (b) Calculated Strehl ratios for the metalens, which indicates nearly diffraction limited focusing at different wavelengths.}
\end{figure}

In contrast, as depicted in  Fig.~\ref{Figure5}(a), the focal length of the achromatic metalens remains largely consistent. The maximum shift in the focal length of the achromatic metalens is only 4$\%$ at 635 nm, demonstrating excellent achromatic performance across the visible light spectrum. To further exemplify polarization insensitivity of achromatic metalens, we simulated the focal spot distribution under various incident polarizations, as detailed in the Section 3 of the Supplementary Materials. The results demonstrate that the metalens can maintain excellent achromatic performance under various polarization incidences. Fig.~\ref{Figure5}(a) also illustrates the focusing efficiency of the achromatic metalens, a round aperture with the diameter three times of the full width at half maximum (FWHM) is defined as the focal region and then the ratio of the intensity integral in the focal region and the region over the entire metalens is calculated as the focusing efficiency \cite{38,60,61}. In the 450-650 nm wavelength range, the focusing efficiency of the metalens ranges from 76.2$\%$ to 85.8$\%$, with an average efficiency of 80.5$\%$. This represents a notable improvement compared to previous designs \cite{55,62}.To illustrate the contributions of co-polarization and cross-polarization to focusing, we measure the electric field components of both at the focal plane across three design wavelengths, detailed in Supplementary Materials Section 4. It is worth noting that the polarization conservation term in the transmitted light reaches its maximum contribution at 450 nm, accounting for approximately 30$\%$ of the energy. This underscores the indispensable role of the co-polarization term in designing efficient broadband achromatic metalenses. Diffraction-limited focal spots are critical for high-performance imaging systems. We simulated the FWHM of the focal spots of the metalens at sampled wavelengths and compared them with the theoretical limit $[k/(2{NA})]$. As depicted in Fig.~\ref{Figure5}(b), the results indicate that Strehl ratios is greater than 80$\%$, demonstrating that the focal spots are at or near the diffraction limit across the entire operational wavelength range of the metalens.

To demonstrate the significance of controlling the polarization conservation component, we design a multi-wavelength polarization-insensitive bi-focus metalens using the principles and noninterlaced structure library described above (the detailed design of the metalens is provided in the Section5, Supplementary Materials).
\begin{figure}
\includegraphics[width=\linewidth]{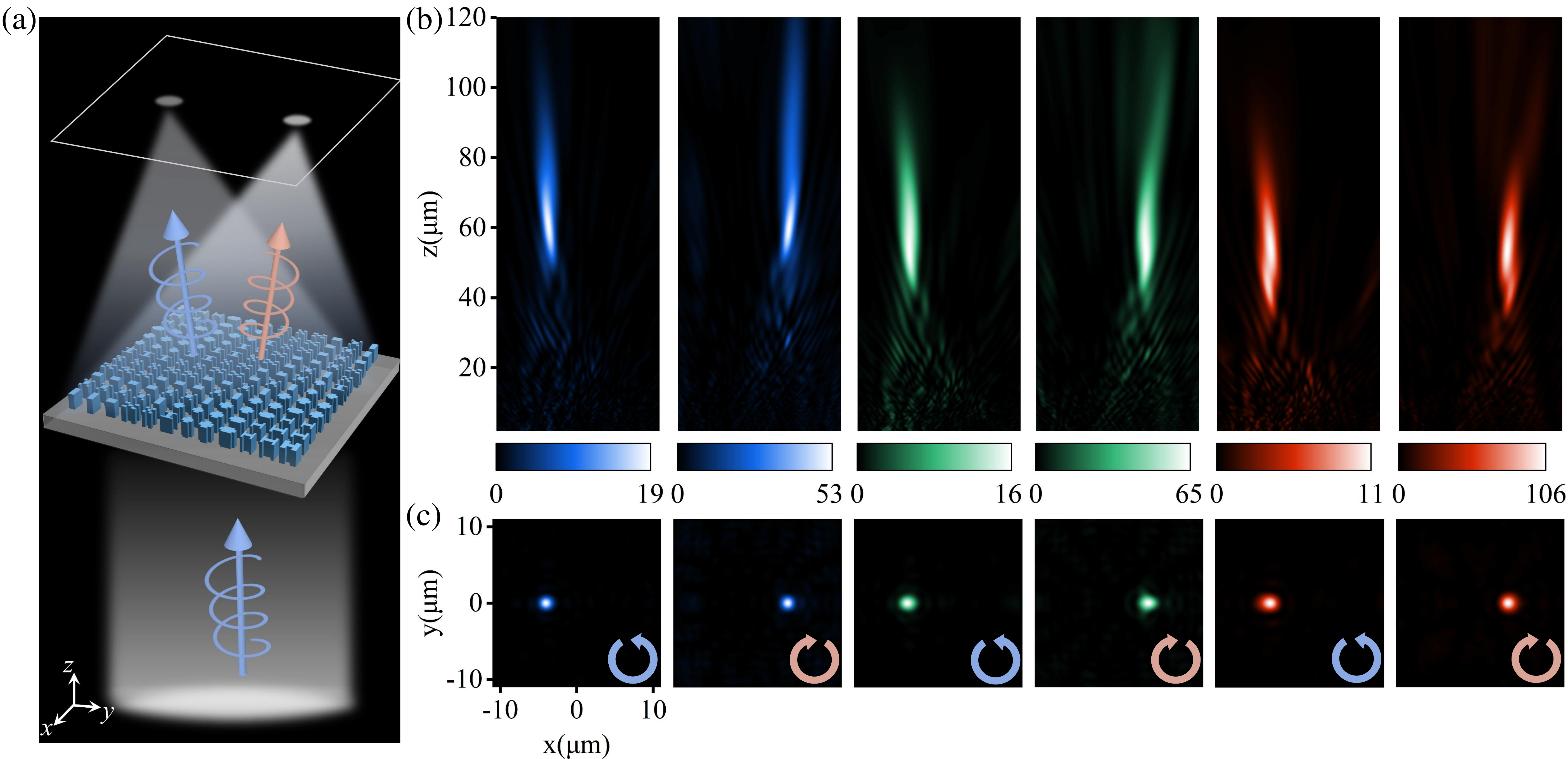}
\caption{\label{Figure6}Working principle and focal distribution of multi-wavelength bi-focal metalens. (a) Under LCP incidence, the multi-wavelength bi-focal metalens focuses the co-polarization and cross-polarization component of the transmitted light at distinct points on the same plane. (b) The intensity distribution of the transmitted light on the \textit{x-z} plane. (c) The intensity distribution on the \textit{x-y} plane at the focus. The blue and red circular arrows represent LCP and RCP component, respectively. The colors blue, green, and red correspond to the wavelengths 450, 532, and 635 nm, respectively.}
\end{figure}
As depicted in Fig.~\ref{Figure6}(a), under LCP incidence, the multi-wavelength bi-focal lens focuses the co-polarization and cross-polarization components of the transmitted light at two distinct focal points on the transverse plane. Fig.~\ref{Figure6}(b) shows the electric field intensity distribution of the LCP/RCP components of the transmitted light in the \textit{x-z} plane. It demonstrates that the metalens exhibits excellent multi-wavelength focusing characteristics. As shown in Fig.~\ref{Figure6}(c), the focal quality of the two polarization components of the transmitted light is high, with almost no crosstalk. Due to the orthogonality of the polarization channels, the metalens can achieve dual focal points in the transverse plane, while maintaining negligible crosstalk between the focal points. According to our previously derived theory, the multi-wavelength bi-focal metalens is polarization-insensitive, capable of accommodating arbitrary polarization incidence. Furthermore, owing to the broadband characteristics of the device, arbitrary dispersion can be achieved by coupling two polarization channels and three wavelength channels. This design strategy is highly significant for applications such as circularly polarized light acquisition \cite{63}, virtual reality \cite{64}, augmented reality \cite{65}, and tomography \cite{66,67}.

\section{Conclusions}

In summary, we proposed a design strategy for an efficient broadband metalens operating in the visible. By maximizing the role of the co-polarization component in enhancing focusing efficiency, rather than considering it as stray light, we curated a minimalist anisotropic nanofin library. Employing the particle swarm algorithm, we optimized the phase distribution of the achromatic metalens at each designed wavelength, effectively reducing phase compensation errors to achieve optimal performance. This approach successfully achieved polarization-independent high-efficiency achromatic focusing across nearly the entire visible light spectrum. Furthermore, we developed a multi-wavelength bi-focal metalens. This design provides considerable flexibility for creating diverse and efficient optical components, making it well-suited for applications such as AR/VR displays and spectral imaging.

\begin{acknowledgments}

This work was supported by the National Natural Science Foundation of China (Grants Nos. 12064025, 12304420, 12264028, 12364045), the Natural Science Foundation of Jiangxi Province (Grants Nos. 20212ACB202006, 20232BAB201040, 20232BAB211025), the Young Elite Scientists Sponsorship Program by JXAST (Grant No.2023QT11).
\end{acknowledgments}

\nocite{*}


%

\end{document}